\documentstyle[multicol,aps,pre,epsf]{revtex}
\begin{document} 
\title{Fraction  of uninfected  walkers in  the one-dimensional  Potts
model}
\author{S. J. O'Donoghue and A. J. Bray}
\address{Department  of  Physics  and  Astronomy,  The  University  of
Manchester, M13 9PL, United Kingdom}
\date{\today}
\maketitle

\begin{abstract}

The dynamics of the one-dimensional $q$-state Potts model, in the zero
temperature  limit, can  be formulated  through the  motion of  random
walkers  which  either  annihilate ($A+A  \rightarrow  \emptyset$)  or
coalesce ($A+A  \rightarrow A$)  with a $q$-dependent  probability. We
consider all of  the walkers in this model to  be mutually infectious.
Whenever  two  walkers  meet, they  experience  mutual  contamination.
Walkers which avoid an encounter with another random walker up to time
$t$ remain uninfected. The fraction  of uninfected walkers is known to
obey a  power law  decay $U(t) \sim  t^{-\phi(q)}$, with  a nontrivial
exponent $\phi(q)$ [C.\ Monthus, Phys.\  Rev.\ E {\bf54}, 4844 (1996);
S.\ N.\  Majumdar and S.\  J.\ Cornell, \textit{ibid.} {\bf  57}, 3757
(1998)].  We probe  the  numerical  values of  $\phi(q)$  to a  higher
degree of accuracy  than previous simulations and  relate the exponent
$\phi(q)$ to  the persistence  exponent $\theta(q)$ [B.\  Derrida, V.\
Hakim  and V.\  Pasquier, Phys.\  Rev.\ Lett.\  {\bf75}, 751  (1995)],
through the relation $\phi(q)=\gamma(q)\theta(q)$  where $\gamma$ is a
new  exponent introduced  in [S.\  J.\  O'Donoghue and  A.\ J.\  Bray,
cond-mat/0111133].  Our  study  is  extended to  include  the  coupled
diffusion-limited reaction $A+A \rightarrow B$, $B+B \rightarrow A$ in
one-dimension with equal  initial densities of $A$  and $B$ particles.
We find that  the density of walkers decays in  this model as $\rho(t)
\sim t^{-1/2}$. The fraction of sites  unvisited by either an $A$ or a
$B$ particle  is found to  obey a  power law, $P(t)  \sim t^{-\theta}$
with  $\theta \simeq  1.33$.  We discuss  these  exponents within  the
context of  the $q$-state Potts  model and present  numerical evidence
that the fraction  of walkers which remain uninfected  decays as $U(t)
\sim  t^{-\phi}$,  where  $\phi  \simeq 1.13$  when  infection  occurs
between  like particles  only, and  $\phi  \simeq 1.93$  when we  also
include cross-species contamination. We find that the relation between
$\phi$ and  $\theta$ in  this model  can also  be characterized  by an
exponent $\gamma$, where similarly, $\phi=\gamma\theta$.

\end{abstract}

\begin{multicols}{2}

\section{Introduction}

When  a  system  is   quenched  from  a  homogeneous  high-temperature
disordered state  into a low-temperature regime,  well defined domains
of equilibrium  ordered phases form  randomly and then grow  with time
(``coarsen'') in  a self-similar  way, until  the domain  size becomes
comparable  to the  size of  the  system \cite{1}.  A dynamic  scaling
hypothesis suggests that,  at late times, the evolution  of the system
is characterized  by a  single length scale  $L(t)$ that  represents a
typical linear size  of the domains. It is well  established, at least
for systems with a scalar order parameter, that $L(t) \sim t^{n}$ with
$n  = 1/2$  for  \textit{non-conserved}  dynamics and  $n  = 1/3$  for
\textit{conserved} dynamics  \cite{1}. The  Ising spin  model evolving
with  Glauber  spin  flip   dynamics  \cite{2}  demonstrates  behavior
characteristic  of the  former while  evolution according  to Kawasaki
spin dynamics \cite{3} exemplifies the latter. A generalization of the
Glauber-Ising  model is  the $q$-state  Potts model  \cite{4,5,6,7,8},
which has  $q$ distinct, but equivalent,  ordered phases. Experimental
realizations  are known  for $q  =  2$ i.e.  the Ising  model and  for
$q=3,4,\infty$  \cite{4}.  The $q  =  \infty$  case describes  several
cellular structures \cite{9}, e.g. polycrystals \cite{10}, soap froths
\cite{11}, magnetic bubbles \cite{12} and foams \cite{13}.

Such  coarsening  systems  are   amongst  those  which  have  received
considerable  attention   in  recent   years  with  regard   to  their
\textit{persistence} properties
\cite{13,14,15,16,17,18,19,20,21,22,23,25,26,27,28,29,30,31,32,33,34,n1,35,36,37,38,39,40,41,42,42a,43,44,45,n2,46}.
The  persistence probability  $P(t)$  is simply  the probability  that
a  given  stochastic  variable   $\phi(x,t)$,  with  zero  mean,  does
not  change  sign  in  the  time  interval  $[0,t]$.  Theoretical  and
computational  studies  of persistence  include  spin  systems in  one
\cite{15,16}  and higher  \cite{17,18,19,20,21} dimensions,  diffusion
fields \cite{22,23}, fluctuating  interfaces \cite{25}, phase-ordering
dynamics    \cite{26,27,28},   Lotka-Volterra    models   \cite{29,30}
and   reaction-diffusion  systems   \cite{31,32,33,34,n1,35,36,37,38}.
Experimental  studies  include  the   coarsening  dynamics  of  breath
figures \cite{39}, foams \cite{13},  soap froths \cite{40,41}, twisted
nematic liquid  crystals \cite{42}, and one-dimensional  gas diffusion
\cite{42a}. Persistence in non-equilibrium critical phenomena has also
been studied in the context of the global order parameter $M(t)$ (e.g.
the total  magnetization of a  ferromagnet), regarded as  a stochastic
process  \cite{43,44}.  In  many  systems of  physical  interest,  the
persistence decays algebraically according to $P(t) \sim t^{-\theta}$,
where  $\theta$  is  the  persistence exponent  and  is,  in  general,
nontrivial. The non-triviality of $\theta$ emerges as a consequence of
the coupling  of the  field $\phi(x,t)$ to  its neighbors,  since such
coupling implies that the stochastic process at a fixed point in space
and time is non-Markovian. For the $q$-state Potts model at $T=0$, the
fraction of  spins which have never  flipped up to time  $t$ (i.e that
fraction that  have \textit{persisted}  in their original  phase), has
been observed \cite{15,17,18,19} to obey  a power law decay $P(t) \sim
t^{-\theta(q)}$,  where  $\theta(q)$  has been  obtained  exactly,  in
one-dimension (1D), by Derrida
\textit{et al.} \cite{16} for arbitrary $q$,
\begin{equation}
\theta(q) = -\frac{1}{8} + \frac{2}{\pi^2}\,
\left[\cos^{-1}\left(\frac{2-q}{\sqrt{2}\,q}\right)\right]^2.
\label{Eq1} 
\end{equation}
There is  a very  direct way \cite{47,48,49}  of relating  the Glauber
dynamics  of the  1D  $q$-state  Potts model  at  zero temperature  to
reaction-diffusion models \cite{50,51}.  Consider uncorrelated initial
conditions where each of the $q$  phases is present with equal density
$1/q$.  During any  time interval  $dt$,  each spin  $S_{i}(t)$ has  a
probability $dt/2$ of becoming equal  to its right neighbor, $dt/2$ of
becoming  equal to  its  left  neighbor and  a  probability $1-dt$  of
retaining its  own value.  The domain  walls therefore  perform random
walks. Upon  contact, two  domain walls react  instantaneously, either
\textit{annihilating}  $A+A  \rightarrow \emptyset$  with  probability
$1/(q-1)$ or \textit{coalescing} $A+A  \rightarrow A$ with probability
$(q-2)/(q-1)$, these  numbers being the probabilities  that the states
on  the far  sides  of  the walkers  are  the  same (annihilation)  or
different (coalescence). This  reaction-diffusion model, together with
the fact  that in  the initial  condition each bond  is occupied  by a
domain wall  with probability  $(q-1)/q$, is completely  equivalent to
the spin problem. The fraction of spins which have never flipped up to
time  $t$ is  then  simply  the fraction  of  sites  which have  never
been  visited by  a  random walker.  The  probabilistic algorithm  for
implementing the  Potts model through the  annihilation or coalescence
of random walkers allows $q$ to be  a real number but is restricted to
$q \ge  2$. However,  an equivalent Ising  spin representation  of the
Potts  persistence  problem \cite{45}  permits  a  study \cite{37}  of
$\theta(q)$ in the regime $1 < q < 2$.

In the  present work  we study  the fraction  of random  walkers which
have never  encountered another  walker, using  the reaction-diffusion
representation of the  1D Glauber-Potts model at  zero temperature. To
facilitate our discussion, we consider  all the random walkers at time
$t = 0$ to be uninfected.  All the walkers are, however, considered to
be  infectious, so  that  for  $t >  0$  any  contact between  walkers
(assuming they  survive the encounter) leads  to mutual contamination.
Walkers that avoid all contact remain uninfected. Our goal, therefore,
is to  address the fraction  of uninfected  walkers $U(t)$ up  to time
$t$.

This  problem  has been  addressed  in  some  detail  by a  number  of
different  authors  \cite{27,n1,n2},  although   under  the  guise  of
a  different  interpretation.  The   general  approach  used  involves
monitoring  the motion  of a  tagged  test particle,  released in  the
system at time  $t=0$. If the tagged particle  diffuses with diffusion
constant  $D'$ and  the other  particles  in the  system diffuse  with
diffusion constant $D$,  the tagged particle is viewed  as an external
impurity  diffusing  through a  homogeneous  background.  One is  then
interested  in  addressing  the  survival probability  of  the  tagged
particle i.e.,  the probability  is has not  been absorbed  by another
particle in the system. Of course in our terminology, this corresponds
to the probability that the  tagged particle remains uninfected. It is
well established \cite{n1,n2} that $U(t)$  decays according to a power
law $U(t)  \sim t^{-\phi(q,c)}$,  where $c=D'/D$ is  the ratio  of the
two  diffusion constants  in the  problem. When  $D'=0$, corresponding
to  a  static  impurity,  the  model clearly  maps  onto  the  problem
of  persistent  spins and  in  this  case, $\phi(q,0)=\theta(q)$.  The
exponent $\phi(q,c)$  is therefore considered a  generalization of the
standard persistence exponent $\theta(q)$, characterizing the survival
probability of a mobile particle  under the coarsening dynamics of the
Potts model.  A recent study of  the survival probability of  a mobile
particle,  moving  according  to either  deterministic  or  stochastic
rules, in given  in \cite{n2} , where the  external fluctuating field
takes  the form  of either  the  solution to  the diffusion  equation,
the  coarsening dynamics  of the  $q$-state Potts  model or  spatially
uncorrelated Brownian signals.

In the limit  $q \rightarrow \infty$ the Potts model  is equivalent to
the  $A+A  \rightarrow  A$  single species  coalescence  process.  The
dynamics of this  system are particularly simple,  since each particle
sees only  the ``cage''  formed by  its two  nearest neighbors.  It is
immaterial  that  these  neighbors  may coalesce  with  more  distant
particles. Consequently,  the kinetics of each  particle involves only
itself and  its two nearest  neighbors and  one is therefore  able to
calculate $\phi(\infty,c)$ exactly \cite{27,n1}:
\begin{equation}
\phi(\infty,c) = \frac{\pi}{ 2\cos^{-1} \left[ \frac{c}{1+c} \right]}\ .
\label{inf}
\end{equation}
(Note that for $c=0$, this of course reduces to the static persistence
exponent,  $\phi(\infty,0)=\theta(\infty)=1$.)  In  the limit  of  the
Ising  model  ($q=2$),  there  is,  however,  no  corresponding  exact
result  for general  $c$.  An exact  result is  only  known for  $c=0$
i.e., $\phi(2,0)=\theta(2)=3/8$.  A mean-field  Smoluchowski \cite{27}
approach  predicts   for  $q=2$  and  general   $c$  that,  $\phi(2,c)
=  \sqrt{(1+c)/8}$,  in  good  agreement  with  numerical  simulations
\cite{27}.  However,  when  extended  to the  large  $q$  limit,  this
Smoluchowski  approach   differs  substantially  from  the   exact  $q
\rightarrow \infty$ result Eq.\ (\ref{inf}). More recently, however, a
perturbation theory  developed by Monthus \cite{n1}  has evaluated the
exponent $\phi(q,c)$ to first order in $q-1$ for arbitrary
$c$ and at first order in $c$ for arbitrary $q$. In this paper, we will
not study the exponent $\phi(q,c)$ in complete generality since we are
primarily interested  in the special  case where all of  the particles
are  equally  mobile  $(D'=D)$  and  henceforth   we  therefore  write
$\phi(q,1)=\phi(q)$.

Within  the context  of  the  current work,  where  we are  interested
in  the  case  $c=1$,  the  random walkers  represent  the  motion  of
domain walls  undergoing zero temperature coarsening.  The fraction of
uncollided domain walls (i.e., uninfected walkers) has previously been
interpreted as ``domain wall persistence'' or equally, the probability
that two adjacent domains survive. Under this interpretation, Majumdar
\textit{et al.} \cite{n2} have studied the quantity $U(t)$ numerically
for  various  $q$.  They  observed  a  power  law  decay,  $U(t)  \sim
t^{-\phi(q)}$ in agreement  with the results of  Monthus \cite{n1} and
obtained a spectrum of values for $\phi(q)$: $1/2 \le \phi(q) \le 3/2$,
where $2 \le q \le \infty$. In this paper, we extend and improve these 
numerical results by  presenting (Sec.\ II)  more accurate  data  than 
previously achieved \cite{n2} and exploring a larger range of $q$.  We 
also relate the value of  $\phi(q)$ to the static  persistent exponent 
$\theta(q)$ via  $\phi(q)=\gamma(q)\theta(q)$, where $\gamma$ is a new  
exponent that we recently introduced in \cite{57}.

We additionally consider (Sec.\ III) the coupled diffusion-limited reaction
$A+A \rightarrow B$,  $B+B \rightarrow A$ in  one-dimension with equal
initial  densities,  $\rho_{A}(0) =  \rho_{B}(0)$.  We  find that  the
density  of particles  decays in  this model  according to  $\rho \sim
t^{-1/2}$, independent  of the initial walker  density, characteristic
of ``Potts like'' behavior \cite{52}.  The fraction of sites unvisited
by either an  $A$ or a $B$ particle decays  as $P(t) \sim t^{-\theta}$
with  $\theta \simeq  1.33$.  We discuss  these  exponents within  the
context of the $q$-state Potts model  and observe that the fraction of
walkers which  remain uninfected by their  \textit{own} species decays
according to $U(t) \sim t^{-\phi}$,  where $\phi \simeq 1.13$. We also
study the fraction of walkers which remain uninfected by \textit{both}
species and  find that this  too obey a  power law, with  $\phi \simeq
1.93$. In analogy to Sec.\ II  we express the exponent $\phi$ is terms
of  the  static persistence  exponent  $\theta$  and similarly  obtain
$\phi=\gamma\theta$. We  conclude in  Sec.\ IV with  a summary  of the
results and a discussion of some open questions.

\section{THE POTTS MODEL}

We investigate  numerically the fraction of  uninfected walkers $U(t)$
in the 1D $q$-state Potts model, at zero temperature, as a function of
$q$. Our  simulations are performed on  a 1D lattice of  size $10^{7}$
with  periodic boundary  conditions. At  $t =  0$ a  random walker  is
placed at  each lattice site,  so that $\rho(0)  = 1$. We  choose this
initial high density  of walkers to accelerate  the system's evolution
into  the asymptotic  regime. Our  model is  updated using  the direct
method  \cite{53} i.e.\  at  each computational  step,  a particle  is
picked at random  and moved with probability $D=1/2$  to a neighboring
site, where $D$ is the diffusion  constant. If the destination site is
occupied,  the  two  particles  either  annihilate,  with  probability
$1/(q-1)$, or coalesce, to become  a single particle, with probability
$(q-2)/(q-1)$, in  accordance with the reaction-diffusion  dynamics of
the Potts model detailed in Sec.\ I. For each move of a particle, time
$t$ is  increased by $dt  = 1/N$, where $N$  is the current  number of
particles  in  the system.  Such  sequential  dynamics can  be  chosen
without  loss  of  generality  as parallel  dynamics  exhibit  similar
asymptotic  behavior  \cite{54}.  Our simulations  are  performed  for
$5000$ time steps and our results are averaged over $100$ runs.

\begin{figure}
\narrowtext
\centerline{\epsfxsize\columnwidth\epsfbox{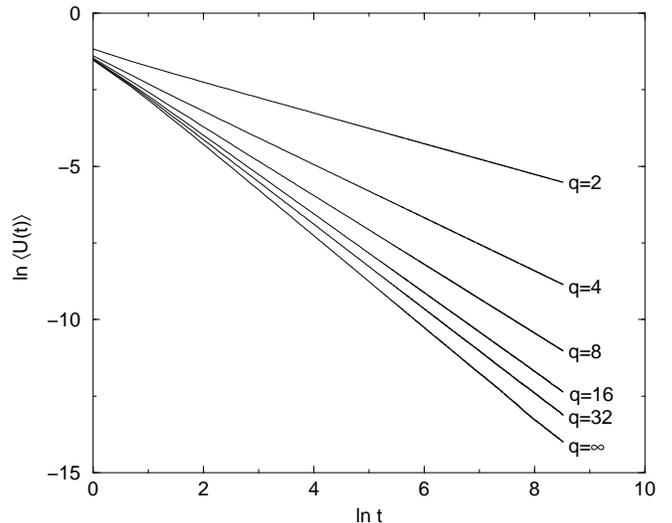}}
\caption{Log-log plot of the fraction of uninfected walkers in the 1D,
$T=0$, Glauber-Potts model as a function of time, for various $q$.}
\label{Fig1}
\end{figure}

\begin{figure}
\narrowtext 
\centerline{\epsfxsize\columnwidth\epsfbox{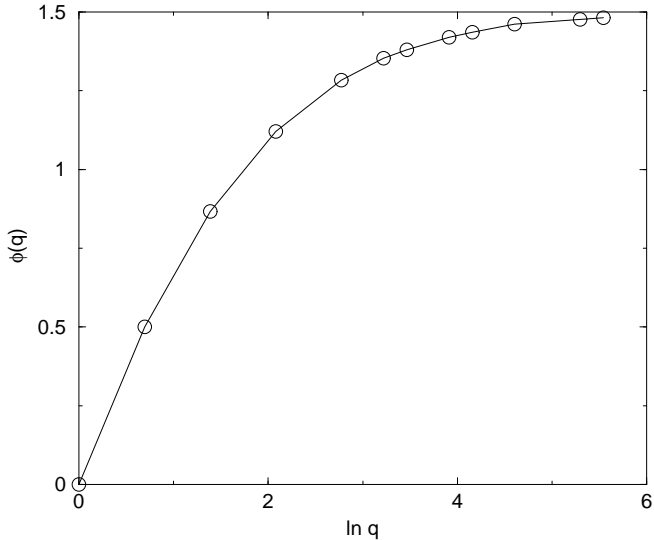}} 
\caption{Linear-log plot  of $\phi$ as a  function of $q$ for  the 1D,
$T=0$, Glauber-Potts model.}
\label{Fig2}
\end{figure}

Fig.\ \ref{Fig1} clearly shows that the fraction of uninfected walkers
decays according to a power-law, $U(t) \sim t^{-\phi(q)}$. In Table\ I
we present  our numerical values for  $\phi(q)$ in the range  $2 \le q
\le  \infty$.  These  values  were obtained  by  performing  a  linear
regression  on  log-log  curves  such  as  those  presented  in  Fig.\
\ref{Fig1}. The regression was taken in the range $10 \le t \le 1000$.
This  range  was chosen  to  avoid  initial  transients and  to  avoid
statistical fluctuations between different runs which become prominent
as $t$  increases, especially  for large $q$.  In Fig.\  \ref{Fig2} we
plot $\phi$  as a function  of $q$, where  we choose to  represent the
data on a linear-log plot merely for clarity of presentation.

\begin{center}
\begin{tabular}{|c|c|}
\hline \hline
\hspace{0.5cm} $q$ \hspace{0.5cm} & \hspace{0.5cm} $\phi$
\hspace{0.5cm} \\ \hline
\hspace{0.5cm} 2 \hspace{0.5cm} & \hspace{0.5cm} 0.5006(7)
\hspace{0.5cm} \\
\hspace{0.5cm} 4 \hspace{0.5cm} & \hspace{0.5cm} 0.867(5)
\hspace{0.5cm} \\
\hspace{0.5cm} 8 \hspace{0.5cm} & \hspace{0.5cm} 1.121(5)
\hspace{0.5cm} \\
\hspace{0.5cm} 16 \hspace{0.5cm} & \hspace{0.5cm} 1.283(6)
\hspace{0.5cm} \\
\hspace{0.5cm} 25 \hspace{0.5cm} & \hspace{0.5cm} 1.352(6)
\hspace{0.5cm} \\
\hspace{0.5cm} 32 \hspace{0.5cm} & \hspace{0.5cm} 1.380(7)
\hspace{0.5cm} \\
\hspace{0.5cm} 50 \hspace{0.5cm} & \hspace{0.5cm} 1.419(7)
\hspace{0.5cm} \\
\hspace{0.5cm} 64 \hspace{0.5cm} & \hspace{0.5cm} 1.435(8)
\hspace{0.5cm} \\
\hspace{0.5cm} 100 \hspace{0.5cm} & \hspace{0.5cm} 1.461(8)
\hspace{0.5cm} \\
\hspace{0.5cm} 128 \hspace{0.5cm} & \hspace{0.5cm} 1.464(8)
\hspace{0.5cm} \\
\hspace{0.5cm} 256 \hspace{0.5cm} & \hspace{0.5cm} 1.481(9)
\hspace{0.5cm} \\
\hspace{0.5cm} $\infty$ \hspace{0.5cm} & \hspace{0.5cm} 1.495(9)
\hspace{0.5cm} \\ \hline \hline
\end{tabular}
\end{center}

\begin{small}
TABLE\ I. Numerical values of the exponent $\phi(q)$ in the 1D, $T=0$,
Glauber-Potts model where $U(t) \sim t^{-\phi(q)}$, for various values
of q.
\end{small}

The case $q = 2$, which corresponds to the Ising model, reduces to the
$A+A  \rightarrow  \emptyset$  reaction-diffusion process,  where  the
density  of  uninfected  walkers  is clearly  equal  to  the  particle
density. It is well known \cite{52}  that in the $q$-state Potts model
the particle density obeys a power-law decay, $\rho(t) \sim t^{-1/2}$,
which is consistent  with our value of $\phi(2) =  0.5006(7)$. For the
case $q  = \infty$, one can  obtain the exact value  of $\phi(\infty)$
from  Eq.\  (\ref{inf})  i.e.   $\phi(\infty,1)=3/2$.  This  is  again
consistent with  our numerical result,  $\phi(\infty)=1.495(9)$. These
limiting values, along with our numerics, suggest that the fraction of
uninfected walkers decays according  to $U(t) \sim t^{-\phi(q)}$ where
$1/2  \le \phi(q)  \le  3/2$, for  $2 \le  q  \le \infty$,  consistent
with the  results of  Monthus \cite{n1}  and Majumdar  \textit{et al.}
\cite{n2}. Although the values of $\phi(q)=1/2$ and $\phi(\infty)=3/2$
at the  limit of the $q$  spectrum are understood, it  was pointed out
by  Monthus  \cite{n1}  that  the  formalism that  led  to  the  exact
determination of the static persistence exponent $\theta(q)$ \cite{16}
cannot be easily applied to compute the exact value of $\phi(q,c)$ for
general  $c$, including  the  limit $c=1$.  The  exact calculation  of
$\phi(q)$ therefore remains an open problem.

\begin{figure}
\narrowtext
\centerline{\epsfxsize\columnwidth\epsfbox{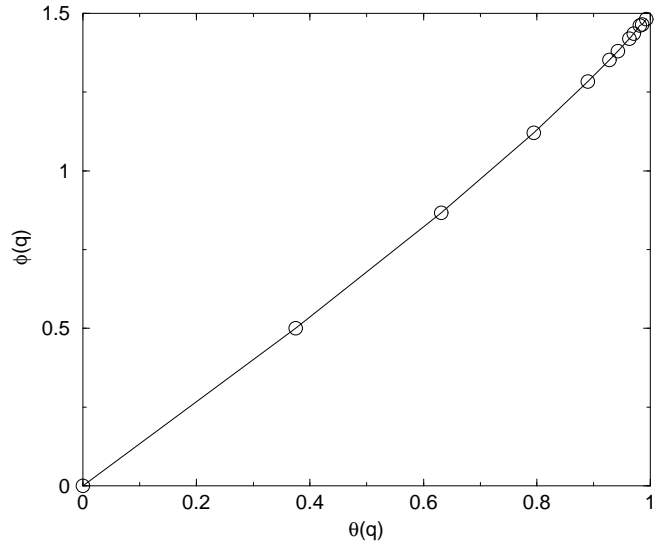}}
\caption{The exponent  $\phi$ plotted  against the  exponent $\theta$,
for various $q$ for the 1D, T=0, Glauber-Potts Model.}
\label{Fig3}
\end{figure}

Recently, we  have explored the connection  between uninfected walkers
and  unvisited  sites  in  one-dimension   within  the  context  of  a
system of  non-interacting randomly  diffusing particles  $A \emptyset
\longleftrightarrow \emptyset A$, and  the $A+B \rightarrow \emptyset$
diffusion-limited reaction \cite{57}.  We found that in  both of these
models,
\begin{equation}
U(t) \simeq [P(t)]^\gamma, \qquad {\rm with}\ \gamma \simeq 1.39.
\label{Eq2}
\end{equation}
While this relationship is obeyed reasonably accurately over the range
of times  accessible to our  simulations, there is evidence,  for both
processes,  that $\gamma$  approaches a  smaller value  at late  times
\cite{57}. Clearly,  we can  write an  analogous relationship  for the
$q$-state  Potts model,  where  $U(t) \sim  t^{-\phi(q)}$, $P(t)  \sim
t^{-\theta(q)}$  and  $\gamma$ is  now  some  function of  $q$,  i.e.\
$t^{-\phi(q)} = t^{-\gamma(q) \theta(q)}$, so that
\begin{equation}
\phi(q) = \gamma(q) \theta(q)\ .
\label{Eq3}
\end{equation}
We  plot  $\phi(q)$ against  $\theta(q)$  in  Fig.\ \ref{Fig3},  where
our  values for  $\phi(q)$  have  been taken  from  Table\  I and  the
corresponding values of $\theta(q)$  have been calculated exactly from
Eq.\ (\ref{Eq1}). The point at  the origin corresponds to $q=1$, where
$\phi$  and  $\theta$ both  vanish.  Fig.\  \ref{Fig3} indicates  that
$\phi(q)$  increases  monotonically  as  a  function  of  $\theta(q)$.
Therefore,  given that  we know  the exact  values of  $\theta(q)$ and
$\phi(q)$ for $q=2,\infty$, and given the convex nature of the plot in
Fig.\ \ref{Fig3},  we can  use Eq.\  (\ref{Eq3}) to  determine precise
bounds on $\gamma(q)$ in this range:
\begin{equation}
4/3 \le \gamma(q) \le 3/2, \qquad q \in [2, \infty].
\label{Eq4}
\end{equation}
Thus  while $\phi(q)$  and $\theta(q)$  are separately  quite strongly
dependent on $q$, their ratio $\gamma(q)$ depends only weakly on $q$.

\section{THE EXCHANGE MODEL}

The  $A+A  \rightarrow  B$,   $B+B  \rightarrow  A$  diffusion-limited
reaction offers  a curious combination  of birth and  death processes.
For brevity, we  dub this model the ``exchange  model''. The particles
in the exchange model diffuse randomly. Whenever a particle encounters
a walker  belonging to its  own species, the two  particles annihilate
instantaneously, to create a  particle belonging to the \textit{other}
species at the point of annihilation.  For the purposes of this study,
we allow  lattice sites  to contain  multiple particles,  enabling two
particles of differing species to  cohabit a single lattice site. This
feature  of our  model  ensures that,  in  one-dimension, the  lattice
structure  $...ABABABABABAB...$  continues  to evolve.  Confining  our
system to a maximum of one particle per site would otherwise mean that
the preceding configuration would be stable for all time $t$. We focus
our attention on the 1D exchange model with equal initial densities of
$A$ and $B$ particles.

Our numerical  simulations are performed on  a one-dimensional lattice
with periodic boundary  conditions. At $t=0$ equal numbers  of $A$ and
$B$ particles are  randomly distributed on the lattice  with a maximum
of one  particle per site.  Our system  then evolves according  to the
dynamics described above. The system is  updated in the same manner as
for the $q$-state Potts model described in Sec.\ II.

\subsection{Particle Density}

\begin{figure}
\narrowtext
\centerline{\epsfxsize\columnwidth\epsfbox{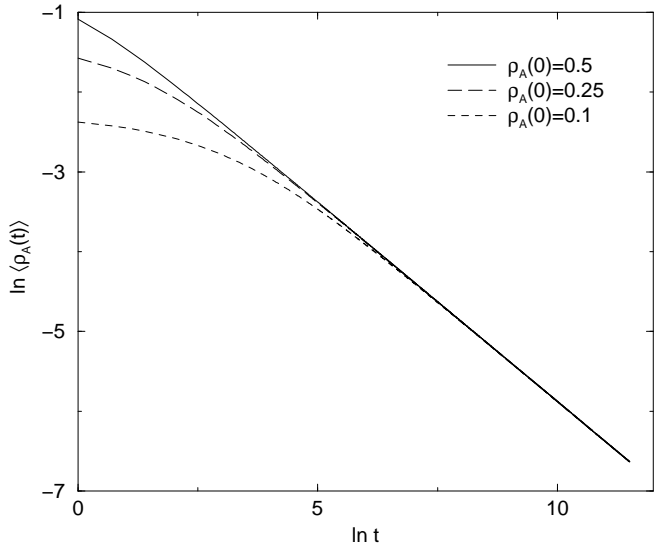}}
\caption{Log-log plot of the particle density as a function of time in
the 1D exchange model.}
\label{Fig4}
\end{figure}

The first question we address  is the particle decay. The unimolecular
reactions  defining  the  exchange  model are  characteristic  of  the
$q$-state Potts model and therefore we can expect the particle density
to  exhibit  similar  asymptotic  behavior.  Indeed,  for  every  pair
of  reactions  between  two  $A$  and  two  $B$  particles,  a  single
$A$  and a  single  $B$ are  produced,  which is  the  same result  as
for  two  $q  =  \infty$  Potts  models,  $A+A  \rightarrow  A$,  $B+B
\rightarrow  B$ operating  on a  single lattice.  We therefore  expect
the  particle density  to decay  in  the exchange  model according  to
$\rho_{A}(t) \sim  Ct^{-\alpha}$, with $\alpha =  1/2$, independent of
the initial  particle density \cite{52}.  This is indeed  confirmed by
our simulations.  In Fig.\ \ref{Fig4}  we present our results  for the
particle density  for three different initial  values of $\rho_{A}(0)$
(recall $\rho_{B}(0)  = \rho_{A}(0)$). The simulations  were performed
on a lattice of size $10^{6}$  for $10^{5}$ time steps and the results
averaged over 100 runs.

In Table\ II we present our  numerical values for the particle density
decay  exponent $\alpha$  and amplitude  $C$, where  $\rho_{A}(t) \sim
Ct^{-\alpha}$.  These  values were  obtained  by  performing a  linear
regression  on the  the plots  in Fig.\  \ref{Fig4} in  the asymptotic
regime, $10^{4} \le t \le 10^{5}$.

\begin{center}
\begin{tabular}{|c|c|c|}
\hline \hline
\hspace{0.4cm} $\rho_{A}(0)$ \hspace{0.4cm} & \hspace{0.4cm} $\alpha$
\hspace{0.4cm} & \hspace{0.4cm} $C$ \hspace{0.4cm} \\ \hline
\hspace{0.4cm} 0.5 \hspace{0.4cm} & \hspace{0.4cm} 0.5007(9)
\hspace{0.4cm} & \hspace{0.4cm} 0.419(5) \hspace{0.4cm} \\
\hspace{0.4cm} 0.25 \hspace{0.4cm} & \hspace{0.4cm} 0.5007(9)
\hspace{0.4cm} & \hspace{0.4cm} 0.419(5) \hspace{0.4cm} \\
\hspace{0.4cm} 0.1 \hspace{0.4cm} & \hspace{0.4cm} 0.5000(5)
\hspace{0.4cm}  &  \hspace{0.4cm}  0.416(9) \hspace{0.4cm}  \\  \hline
\hline
\end{tabular}
\end{center}

\begin{small}
TABLE\ II.  Numerical values for  the particle density  decay exponent
$\alpha$  and  the amplitude  $C$  in  the  1D exchange  model,  where
$\rho_{A}(t)  \sim  Ct^{-\alpha}$,  for  different  initial  densities
$\rho_{A}(0)$.
\end{small}

For the 1D $q$-state Potts model, the amplitude $C$ is $q$-dependent
\cite{52},
\begin{equation}
C(q) = \frac{q-1}{q}\frac{1}{\sqrt{2 \pi D}}.
\label{Eq5}
\end{equation}
Is there  perhaps a  value of  $q$ which  corresponds to  the exchange
model? A value of $q = 4$  in Eq.\ (\ref{Eq5}), with $D=1/2$ returns a
value  $C(4) =  0.423$,  in  good agreement  with  the  values of  $C$
recorded  in Table\  II. Notice  that this  value lies  exactly midway
between  the minimum  value, $C(2)  =  0.282$ and  the maximum  value,
$C(\infty) = 0.564$,  set by the $q$-state Potts  model. Our numerical
results therefore suggest, that in  terms of the particle density, the
two reactions which define the exchange model ($A+A \rightarrow B, B+B
\rightarrow A$) are equivalent, in some loose sense, to two independent
$q=4$ Potts models, though we would not wish to stretch this comparison 
too far. 

\subsection{Fraction of Unvisited Sites}

We  now  turn our  attention  to  the  persistence properties  of  the
exchange  model. Here,  we are  interested in  the fraction  of sites,
$P(t)$,  that have  never  been visited  by  either an  $A$  or a  $B$
particle up to time $t$. An obvious starting point is to see if $P(t)$
obeys a power-law decay with an exponent equal to that returned in the
case  of  two $q  =  4$  Potts Models  i.e.  $P(t)  \sim t^{-2  \times
\theta(q=4)}$ where,  using Eq.\ (\ref{Eq1}), $2  \times \theta(q=4) =
1.263$.  Our  numerical simulations  do  indeed  indicate a  power-law
decay,  $P(t)  \sim  t^{-\theta_{ex}}$. However,  the  decay  exponent
$\theta_{ex} \simeq  1.33$, in disagreement  with what one  might have
naively expected from a study of the particle decay.

\begin{figure}
\narrowtext
\centerline{\epsfxsize\columnwidth\epsfbox{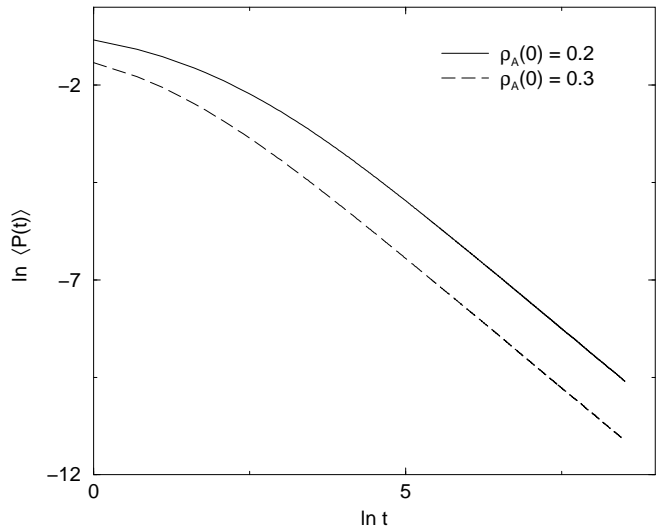}}
\caption{Log-log plot of the fraction of unvisited sites as a function
of time in the 1D exchange model.}
\label{Fig5}
\end{figure}

Our simulations are  performed on a lattice of size  $10^{6}$ for 5000
time steps and the results averaged over 200 runs. In Fig.\ \ref{Fig5}
we present a log-log plot of $P(t)$. In Table\ III we show the results
of a linear regression performed on the curves in Fig.\ \ref{Fig5}, to
obtain values  of $\theta_{ex}$. The  regression was performed  in the
regime $500 \le t \le 5000$, to avoid initial transients.

\begin{center}
\begin{tabular}{|c|c|}
\hline \hline
\hspace{0.5cm}    $\rho_{A}(0)$   \hspace{0.5cm}    &   \hspace{0.5cm}
$\theta_{ex}$ \hspace{0.5cm} \\ \hline
\hspace{0.5cm} 0.2 \hspace{0.5cm} & \hspace{0.5cm} 1.330(7)
\hspace{0.5cm} \\
\hspace{0.5cm} 0.3 \hspace{0.5cm} & \hspace{0.5cm} 1.333(5)
\hspace{0.5cm} \\ \hline \hline
\end{tabular}
\end{center}

\begin{small}
TABLE\  III.  Numerical  values  for the  persistence  decay  exponent
$\theta_{ex}$   in   the  1D   exchange   model,   where  $P(t)   \sim
t^{-\theta_{ex}}$, for initial densities $\rho_{A}(0)$.
\end{small}

\subsection{Fraction of Uninfected Walkers}

Finally,  we  consider  the  fraction of  uninfected  walkers  in  the
exchange  model. We  focus, without  loss of  generality, on  just the
fraction  of uninfected  $A$ particles.  In  this model,  we are  only
interested  in \textit{original}  $A$ particles.  Those $A$  particles
which have been created through the reaction $B+B \rightarrow A$ after
$t=0$, are  also therefore  considered to be  infected. There  are two
distinct  cases  to  consider:  (i)  the  fraction  of  $A$  particles
which remain  uninfected by \textit{only}  their own species  and (ii)
the  fraction  of  $A$  particles  which  have  avoided  infection  by
\textit{both} types of  species. Our approach is  wholly numerical. We
address the former case first.

\begin{figure}
\narrowtext
\centerline{\epsfxsize\columnwidth\epsfbox{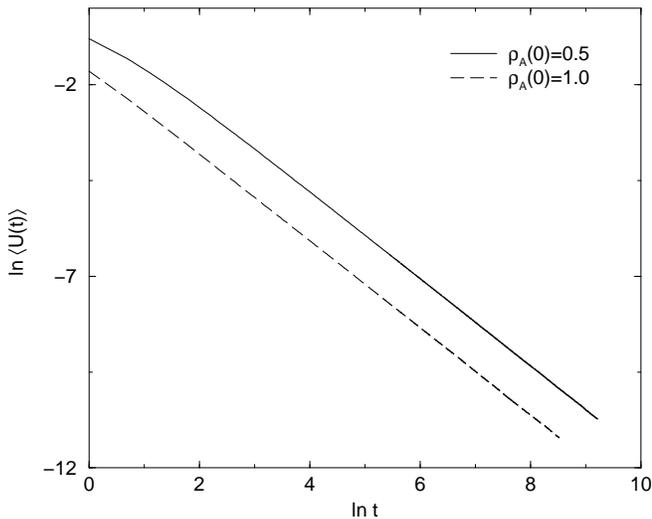}}
\caption{Log-log plot of the fraction of $A$ walkers uninfected by $A$
particles in the 1D exchange model as a function of time.}
\label{Fig6}
\end{figure}

Fig.\ \ref{Fig6} indicates that the fraction of uninfected $A$ walkers
obeys a power-law decay according  to $U(t) \sim t^{-\phi_{A}}$, where
the  subscript $A$  on  the  exponent denotes  infection  by only  $A$
particles (case (i)). The results for $\rho_{A}(0) = 1$ were generated
from 500 runs  on a lattice of  size $5 \times 10^{5}$  over 5000 time
steps.  The results  for $\rho_{A}(0)  = 0.5$  were generated  using a
lattice  of  size $10^{6}$,  run  for  $10^{4}$  time steps  and  also
averaged over 500 runs. In Table\  IV, we present our numerical values
of $\phi_{A}$ for two different initial starting densities. The values
of the exponent were obtained by performing a linear regression on the
curves in Fig.\ \ref{Fig6} in the regime $100 \le t \le 5000$.

\begin{center}
\begin{tabular}{|c|c|}
\hline \hline
\hspace{0.5cm}    $\rho_{A}(0)$   \hspace{0.5cm}    &   \hspace{0.5cm}
$\phi_{A}$ \hspace{0.5cm} \\ \hline
\hspace{0.5cm} 1.0 \hspace{0.5cm} & \hspace{0.5cm} 1.133(5)
\hspace{0.5cm} \\
\hspace{0.5cm} 0.5 \hspace{0.5cm} & \hspace{0.5cm} 1.138(8)
\hspace{0.5cm} \\ \hline \hline
\end{tabular}
\end{center}

\begin{small}
TABLE\  IV. Numerical  values for  the exponent  $\phi_{A}$ in  the 1D
exchange model, where $U(t) \sim t^{-\phi_{A}}$, for initial densities
$\rho_{A}(0)$.
\end{small}

\begin{figure}
\narrowtext
\centerline{\epsfxsize\columnwidth\epsfbox{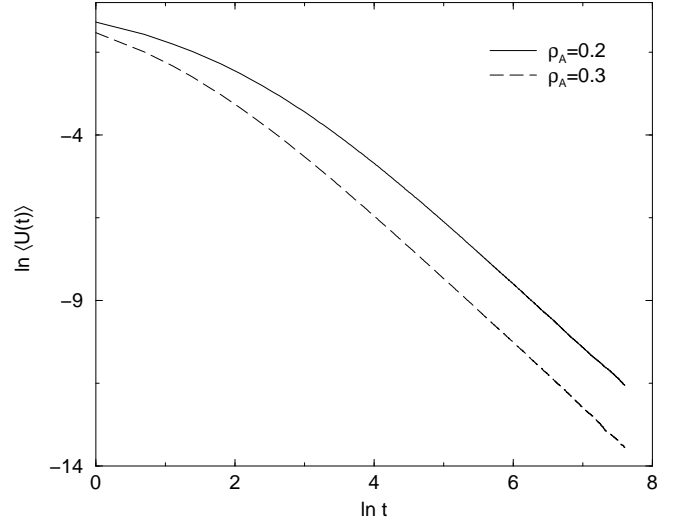}}
\caption{Log-log plot  of the  fraction of  $A$ walkers  uninfected by
both $A$ and $B$  particles in the 1D exchange model  as a function of
time.}
\label{Fig7}
\end{figure}

We now  address case (ii),  the fraction  of $A$ particles  which have
avoided infection  by both  $A$ and $B$  walkers. Our  simulations are
performed on  a lattice  of size  $10^{6}$ for  $2000$ time  steps and
averaged over $500$  runs. In line with our previous  data, we present
our results for $U(t)$ on a log-log plot in Fig.\ \ref{Fig7}.

\begin{figure}
\narrowtext
\centerline{\epsfxsize\columnwidth\epsfbox{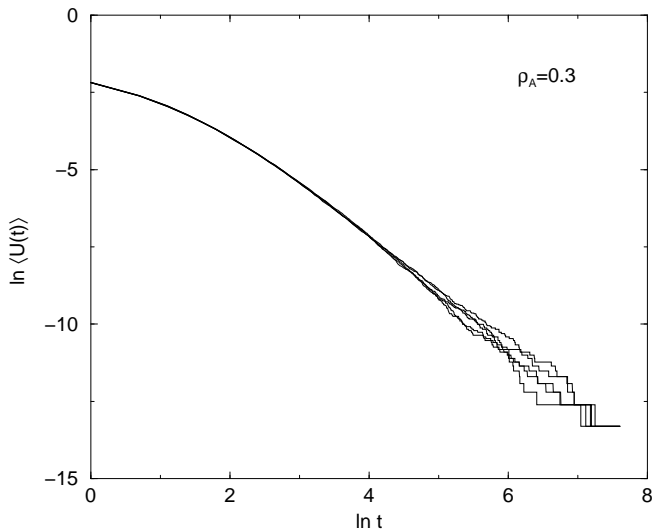}}
\caption{Log-log plot  of the  fraction of  $A$ walkers  uninfected by
both $A$ and $B$  particles in the 1D exchange model  as a function of
time, for five individual simulation runs.}
\label{Fig8}
\end{figure}

\noindent  Unfortunately,  we  do  not obtain  straight  line  graphs.
However,  an  inspection  of  individual   runs  of  the  data  (Fig.\
\ref{Fig8})  shows  that,  in  this  model,  statistical  fluctuations
between different runs become  prominent very quickly, whereas earlier
times are governed by transient effects. The dominance of these two 
regimes renders any determination of an exponent particularly 
difficult.  Using a lower initial density,
with the  intention of delaying the  onset of noisy data  only extends
the  transient  regime, whereas  using  a  higher initial  density  to
minimize transient effects accelerates  the onset of large statistical
fluctuations. In view of the  power-law behavior we identified in case
(i)  and  the  general  trend  of our  data,  we  suggest  that,  with
sufficiently good  statistics, the fraction of  $A$ walkers uninfected
by either  $A$ or $B$ particles  would also exhibit a  power-law decay
$U(t)  \sim  t^{-\phi_{AB}}$,  where  the subscript  on  the  exponent
denotes infection is  permitted by both species. There  is, however, a
cleaner method of extracting the  exponent $\phi_{AB}$. We suggest, in
analogy  with the  other models  we  have studied,  that $U(t)  \simeq
[P(t)]^{\gamma_{ex}}$ so that,
\begin{equation}
\phi_{AB} = \gamma_{ex} \theta_{ex}.
\label{Eq6}
\end{equation}

\begin{figure}
\narrowtext
\centerline{\epsfxsize\columnwidth\epsfbox{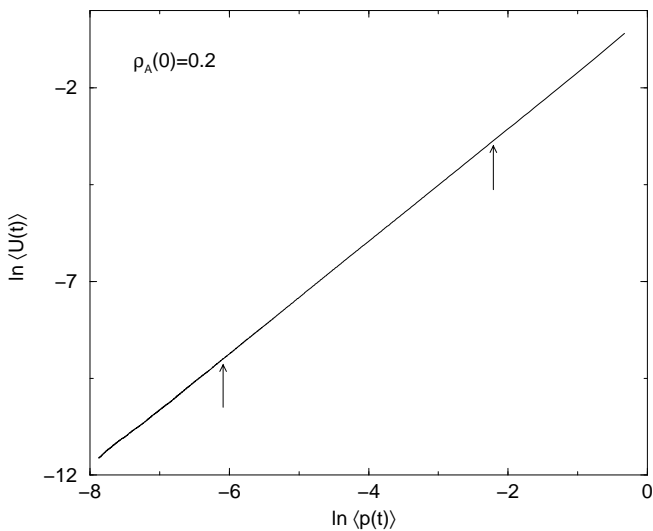}}
\caption{Log-log plot  of the  fraction of  $A$ walkers  uninfected by
both $A$ and  $B$ particles, against the fraction  of unvisited sites,
in the 1D exchange model.}
\label{Fig9}
\end{figure}

If  we  can  establish  the  validity of  the  relation  $U(t)  \simeq
[P(t)]^{\gamma_{ex}}$, and  determine the  value of  $\gamma_{ex}$, we
can then read off the  exponent $\phi_{AB}$ from Eq.\ (\ref{Eq6}). Our
initial condition  specifies a  maximum of one  particle per  site, so
$U(0)=1$. While, in  the continuum description of  the model, $P(0)=1$
also, on a lattice this becomes $P(0)=1-\rho_{0}$. Therefore, to place
these two  quantities on an equal  footing in our numerical  study, we
define, $p(t)=P(t)/P(0)$, so that both  $U(0)$ and $p(0)$ are equal to
unity.  Note,  of  course,  that in  the  continuum  limit,  $\rho_{0}
\rightarrow  0$, we  recover $p(t)  \rightarrow P(t)$.  We present  in
Fig.\ \ref{Fig9} a log-log plot of $U(t) \textit{ vs. } p(t)$.

\noindent  The plot  in Fig.\  \ref{Fig9}  is quite  linear, but  does
not  go through  the  origin,  suggesting that  in  fact $U(t)  \simeq
Kp(t)^{\gamma_{ex}}$ with  $K \ne  1$. We note  that in  the numerical
analysis of  the relationship between noninfectedness  and persistence
in the  $A\emptyset \leftrightarrow \emptyset A$  and $A+B \rightarrow
\emptyset$  models \cite{57}  the equivalent  prefactor $K$  was close
enough to  unity ($K \simeq  0.99$) to  tempt the speculation  that it
might become  unity in  the continuum (low  density) limit,  though we
stress  that  in  the  latter  models there  was  good  evidence  that
some  of the  simulations had  not reached  the asymptotic  regime. In
order  to evaluate  $\gamma_{ex}$ and  $K$ for  the present  model, we
perform a  linear regression on  the data  in Fig.\ \ref{Fig9}  in the
region indicated  by the  arrows, thereby avoiding  initial transients
associated  with  lattice  effects,   and  the  onset  of  statistical
fluctuations  between different  runs of  the data.  We summarize  our
results for $\gamma_{ex}$ and $K$ in Table\ V.

\begin{center}
\begin{tabular}{|c|c|c|}
\hline \hline
\hspace{0.5cm}    $\rho_{A}(0)$   \hspace{0.5cm}    &   \hspace{0.5cm}
$\gamma_{ex}$  \hspace{0.5cm} &  \hspace{0.5cm} $K$  \hspace{0.5cm} \\
\hline
\hspace{0.5cm} 0.2 \hspace{0.5cm} & \hspace{0.5cm} 1.453(7)
\hspace{0.5cm} & \hspace{0.5cm} 0.863(9) \hspace{0.5cm} \\
\hspace{0.5cm} 0.3 \hspace{0.5cm} & \hspace{0.5cm} 1.456(7)
\hspace{0.5cm}  &  \hspace{0.5cm}  0.754(9) \hspace{0.5cm}  \\  \hline
\hline
\end{tabular}
\end{center}

\begin{small}
TABLE\ V. Numerical values of $\gamma_{ex}$ and $K$ in the 1D exchange
model,  where $U(t)  \simeq  K[p(t)]^{\gamma_{ex}}$,  for two  initial
densities $\rho_{A}(0)$.
\end{small}

\begin{figure}
\narrowtext
\centerline{\epsfxsize\columnwidth\epsfbox{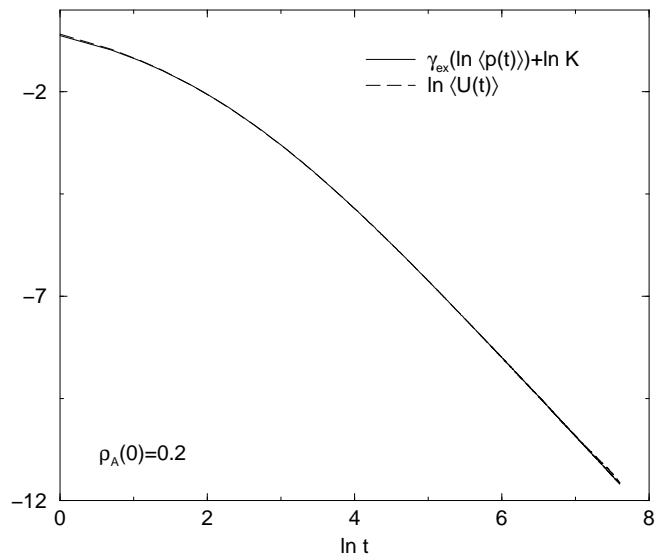}}
\caption{$\ln U(t)$  and $\gamma_{ex} \ln p(t)  + \ln K$ plotted  as a
function of time, to collapse the  data for the uninfected walkers and
unvisited  sites onto  a  single  curve. We  present  our results  for
$\rho_{A}(0)=0.2$ and use $\gamma_{ex}=1.45$ and $K=0.86$.}
\label{Fig10}
\end{figure}

\noindent Note that the value of $K$ for $\rho_{A}(0) = 0.2$ is larger
than that for $\rho_{A}(0) =  0.3$, suggesting the possibility that $K
\rightarrow  1$ as  $\rho_{A}(0) \rightarrow  0$, consistent  with our
results  for  other  models  \cite{57}, though  we  have  no  concrete
arguments to  support this idea.  To test  our results, we  attempt to
collapse the data  for the uninfected fraction and  the persistence on
to a single  curve by plotting $\ln U(t)$ and  $\gamma_{ex} \ln p(t) +
\ln K$ as functions of time $t$, in Fig.\ \ref{Fig10}.

\noindent  The data  collapse  is excellent  over  both transient  and
asymptotic  regimes,   suggesting  that  the  relation   $U(t)  \simeq
K[p(t)]^{\gamma_{ex}}$  holds to  a high  degree of  accuracy for  all
times $t$. We attribute any deviations  between the two curves at late
times to the onset of  statistical fluctuations between different runs
of the  data. Clearly we can  improve the collapse at  either early or
late times by performing the  linear regression in Fig.\ \ref{Fig9} on
the appropriate part of the curve.  We find that over the whole regime
presented  in Fig.\  \ref{Fig10}, $\gamma_{ex}  \simeq 1.45(1)$  works
well. For brevity  and clarity of presentation, we  have presented our
results for $\rho_{A}(0)=0.2$ only,  although we have achieved equally
good results for $\rho_{A}(0)=0.3$.

The  values $\gamma_{ex}  \simeq 1.45$  and $\theta_{ex}  \simeq 1.33$
yield,  using Eq.\  (\ref{Eq6}),  $\phi_{AB} \simeq  1.45 \times  1.33
\simeq 1.93$.  We therefore  argue that, when  infection is  caused by
both species, $U(t) \sim t^{-\phi_{AB}}$ with $\phi_{AB} \simeq 1.93$.

\section{Discussion and Summary}

In  this  paper,  we  have investigated  the  fraction  of  uninfected
walkers  $U(t)$ in  the  1D  $q$-state Potts  model  evolving at  zero
temperature. Our  numerical results are consistent  with previous work
\cite{27,n1,n2},  suggesting  that  $U(t) \sim  t^{-\phi(q)}$  with  a
nontrivial exponent $\phi(q)$, where $1/2  \le \phi(q) \le 3/2$ for $2
\le q  \le \infty$.  Although the  values of  $\phi(q)$ for  $q=2$ and
$q=\infty$ are understood, the exact  calculation of $\phi(q)$ for all
$q$, appears to be a hard problem. In analogy with the other models we
have  studied  \cite{57},  we  have  reduced  the  familiar  study  of
persistent sites $P(t)$, to a limiting case in the study of uninfected
walkers and noted  that $U(t) \sim [P(t)]^{\gamma(q)}$,  with $4/3 \le
\gamma \le  3/2$ for  $2 \le q  \le \infty$. We  discussed in  Sec.\ I
that, with regard to the persistence properties of the Potts model, it
is possible  to probe  the region $1  \le q \le  2$. Although  we have
included a point for the  state $q=1$, where $\theta=\phi=0$, in Fig.\
\ref{Fig2} and Fig.\  \ref{Fig3}, we leave a study of  the fraction of
uninfected walkers in this latter regime as an open challenge.

We have also  studied the 1D exchange model, defined  by the reactions
$A+A \rightarrow  B$, $B+B  \rightarrow A$. Our  numerical simulations
indicate similarities  between this  diffusion-limited system  and the
Potts class of  models, in that the particle  density decays according
to  $\rho(t)  \sim  t^{-1/2}$,  independent of  the  initial  particle
density.  The  amplitude  of  the particle  decay  lies  approximately
halfway between the minimum and maximum  values set by the Potts model
suggesting  that,  within  the  context of  the  particle  decay,  the
exchange model roughly mimics the behavior  of two independent $q = 4$
Potts models. We have also shown  that the fraction of sites unvisited
by either  an $A$  or a  $B$ particle decays  according to  $P(t) \sim
t^{-\theta_{ex}}$ where  $\theta_{ex} \simeq  1.33$. Our study  of the
fraction of  uninfected walkers in  the exchange model focused  on two
cases. In the first instance, we addressed the fraction of $A$ walkers
infected  by only  their own  species  and established  that, in  this
case,  $U(t) \sim  t^{-\phi_{A}}$ where  $\phi_{A} \simeq  1.13$. When
cross species  contamination was also  included in our  simulation, we
found  $U(t) \sim  t^{-\phi_{AB}}$  with $\phi_{AB}  \simeq 1.93$.  We
presented numerical  evidence that  $U(t)\simeq K[p(t)]^{\gamma_{ex}}$
with $\gamma_{ex} \simeq  1.45$. An obvious goal for the  future is to
establish a theoretical framework for  the exponents that occur in the
exchange model  and, in particular,  for the observed  simple relation
between $U(t)$ and $P(t)$. Across the range of models we have studied,
in  both  this and  our  previous  paper  \cite{57}, the  fraction  of
uninfected  walkers,  $U(t)$, and  the  fraction  of unvisited  sites,
$P(t)$,  are  seemingly related  by  $U(t)  \sim [P(t)]^\gamma$.  This
remarkable  relationship, which  seems to  hold reasonably  accurately
across the entire time regime accessible to simulation, merits further
attention. In Table\  VI, we summarize the values of  $\gamma$ for the
various  models we  have studied  (note, however,  that there  is some
evidence  \cite{57} that,  for the  first two  models, the  asymptotic
value of $\gamma$ is smaller 1.39).

\begin{center}
\begin{tabular}{|c|c|}
\hline \hline
\hspace{0.3cm} Model \hspace{0.3cm} & \hspace{0.3cm} $\gamma$
\hspace{0.3cm} \\ \hline
\hspace{0.3cm} $A \emptyset \longleftrightarrow \emptyset A$
\hspace{0.3cm} & \hspace{0.3cm} 1.39 \hspace{0.3cm} \\
\hspace{0.3cm} $A+B \rightarrow \emptyset$ \hspace{0.3cm} &
\hspace{0.3cm} 1.39 \hspace{0.3cm} \\
\hspace{0.3cm} $q$-state  Potts Model \hspace{0.3cm}  & \hspace{0.3cm}
$[4/3, 3/2]$, $q \in [2, \infty]$ \hspace{0.3cm} \\
\hspace{0.3cm} exchange Model \hspace{0.3cm} & \hspace{0.3cm} 1.45
\hspace{0.3cm} \\ \hline \hline
\end{tabular}
\end{center}

\begin{small}
TABLE\ VI. Numerical  results for the exponent $\gamma$  in various 1D
models for which $U(t) \simeq  [P(t)]^{\gamma}$ (results for the first
two models from ref.\ \cite{57}).
\end{small}

\noindent Finally, we note that all values of $\gamma$ obtained so far
lie in the  range $4/3 \le q  \le 3/2$ spanned by  the $q$-state Potts
model  with  $2  \le  q  \le  \infty$.  Up  to  now  our  analysis  of
$\gamma$ has been  purely numerical, and naturally  a more fundamental
understanding of this exponent is required.

\section{ACKNOWLEDGMENT}
We thank Satya Majumdar for drawing our attention to some related work. 
This work was supported by EPSRC (UK).

\end{multicols}

\end{document}